\documentclass[journal]{IEEEtran}
\usepackage{soul,color} % highlight

\usepackage{graphicx}
\usepackage{amsmath}
\usepackage{dcolumn}
\newcolumntype{d}[0]{D{.}{.}{1.2}<{\hspace{-0.0cm}}}
\usepackage{url}
\begin{document}
\title{An Evaluation of Intrusive Instrumental Intelligibility Metrics}

\author{Steven Van Kuyk,~\IEEEmembership{Student Member,~IEEE,}
        W. Bastiaan Kleijn,~\IEEEmembership{Fellow,~IEEE,}
        and Richard C. Hendriks,~\IEEEmembership{Member,~IEEE}% <-this % stops a space
%\thanks{Manuscript received ?; revised ?.}%
\thanks{S. Van Kuyk is with the Victoria University of Wellington, Wellington 6012,
New Zealand (e-mail: steven.van.kuyk@ecs.vuw.ac.nz).}% <-this % stops a space
\thanks{W. B. Kleijn is with the Victoria University of Wellington, Wellington 6012,
New Zealand, and also with the Delft University of Technology, Delft 2628 CD,
The Netherlands (e-mail: bastiaan.kleijn@ecs.vuw.ac.nz).}% <-this % stops a space
\thanks{R. C. Hendriks is with the Delft University of Technology, Delft 2628 CD,
The Netherlands (e-mail: r.c.hendriks@tudelft.nl).}
}


\markboth{ACCEPTED FOR PUBLICATION IN IEEE/ACM TRANSACTIONS ON AUDIO, SPEECH, AND LANGUAGE PROCESSING, DOI: 10.1109/TASLP.2018.2856374}%
%\markboth{DRAFT VERSION}%
{Shell \MakeLowercase{\textit{et al.}}: Bare Demo of IEEEtran.cls for IEEE Journals}



\maketitle

\begin{abstract}%\wbk{I would compress the abstract a bit. It feels too long.}
Instrumental intelligibility metrics are commonly used as an alternative to listening tests. This paper evaluates 12 monaural intrusive intelligibility metrics: SII, HEGP, CSII, HASPI, NCM, QSTI, STOI, ESTOI, MIKNN, SIMI, SIIB, and $\text{sEPSM}^\text{corr}$. In addition, this paper investigates the ability of intelligibility metrics to generalize to new types of distortions and analyzes why the top performing metrics have high performance. The intelligibility data were obtained from 11 listening tests described in the literature. The stimuli included Dutch, Danish, and English speech that was distorted by additive noise, reverberation, competing talkers, pre-processing enhancement, and post-processing enhancement. %STOI, which is arguably the most popular intelligibility metric, achieved a correlation with listening test scores on average of $\rho=0.80$, and its successor, ESTOI, achieved $\rho=0.86$. 
SIIB and HASPI had the highest performance achieving a correlation with listening test scores on average of $\bf \rho=0.92$ and $\bf \rho=0.89$, respectively. %, and a tau coefficient on average of $\bf \tau=0.79$ and $\bf \tau = 0.76$, respectively.
The high performance of SIIB may, in part, be the result of SIIBs developers having access to all the intelligibility data considered in the evaluation. The results show that intelligibility metrics tend to perform poorly on data sets that were not used during their development. By modifying the original implementations of SIIB and STOI, the advantage of reducing statistical dependencies between input features is demonstrated. 
Additionally, the paper presents a new version of SIIB called $\text{SIIB}^\text{Gauss}$, which has similar performance to SIIB and HASPI, but takes less time to compute by two orders of magnitude.

%The high performance of SIIB may be criticized on the grounds that the developers of SIIB were the only researchers with access to all the intelligibility data considered in the evaluation. The results show that many intelligibility metrics perform poorly on data sets that were not used during their development, thus caution should be taken when using intelligibility metrics to replace listening tests, especially for listening conditions where the accuracy of the metric has not been verified.}
\end{abstract}


% \fi
\IEEEpeerreviewmaketitle

%%%%%%%%%%%%%%%%%%%%%%%%%%%%%%%%%%%%%%%%%%%%%%%%
\section{Introduction}

\IEEEPARstart{W}{hen} designing a speech-based communication system it is important to understand how the system will affect the intelligibility and quality of speech. Intelligibility is often defined as the proportion of words correctly identified by a listener \cite{allen2005articulation}, whereas speech quality refers to the pleasantness of the speech signal \cite{loizou2013speech}. Many algorithms for predicting the intelligibility of a communication system have been proposed. This paper summarizes existing algorithms and evaluates their accuracy using data from formal listening tests.%To our knowledge, in terms of the number of listening tests and intelligibility metrics, this is the most comprehensive evaluation of intelligibility metrics to date.

In \cite{shannon1948mathematical}, Shannon proposed that any communication system can be modelled by three components: a transmitter, a receiver, and a channel. In the context of speech communication, the transmitter is the vocal apparatus of the talker, the receiver is the auditory system of the listener, and the channel is the physical medium traversed by the speech signal. The channel may distort the speech signal and decrease the speech signal's intelligibility or quality. As an example, for telephone systems, the speech signal is sampled, quantized, and compressed prior to transmission. Additionally, environmental degradation such as additive noise and reverberation may be introduced at the far-end (i.e., at the talker) or the near-end (i.e., at the listener).

To combat environmental degradation, a variety of speech enhancement algorithms have been proposed (see \cite{loizou2013speech} for an overview). There are two main approaches to speech enhancement: 1) the speech signal can be modified prior to degradation (e.g., optimal energy redistribution \cite{taal2014speech} and dynamic range compression \cite{zorila2012speech}), or 2) the speech signal can be modified after degradation has been introduced (e.g., %spectral subtraction \cite{boll1979suppression} and 
Wiener filters \cite{wiener1949extrapolation}). The former type of algorithm is referred to as a pre-processing algorithm and the latter as a post-processing algorithm. 

%Several experiments have suggested that for single-microphone scenarios, post-processing algorithms are likely to improve speech quality, but are unlikely to significantly improve intelligibility \cite{?}. In contrast, pre-processing algorithms have been known to consistently improve intelligibility \cite{?}. A possible explanation of this trend is given by the data processing inequality (DPI) \cite{?}, which is an information theoretic theorem. The DPI states that if the communication process can be modelled by a Markov chain, then the mutual information between the clean and degraded signal cannot be increased by manipulating the degraded signal.

A key component to the design of speech-based communication systems is an understanding of how they affect intelligibility. Although formal listening tests can provide valid data, such tests are time-consuming, laborious, and expensive. %This makes it infeasible to continuously evaluate a system throughout the design process. 
For this reason, quantities that are fast to compute and correlated with intelligibility are of interest. Such quantities are referred to as {\emph{instrumental intelligibility metrics}}. 

Rather than using human subjects, instrumental intelligibility metrics may rely on knowledge of the clean speech, distorted speech, and the communication channel. There are two types of intelligibility metrics: intrusive and non-intrusive. Intrusive intelligibility metrics require knowledge of the clean speech and either the channel or the distorted speech, whereas non-intrusive intelligibility metrics require only the distorted speech. Although non-intrusive metrics are more widely applicable, they tend to be less correlated with intelligibility than intrusive metrics \cite{falk2015objective, andersen2017non}. From here on, this paper focuses on intrusive intelligibility metrics.

One of the first intelligibility metrics was developed during the 1920's and is called the articulation index (AI) \cite{french1947factors}. The AI %aims to predict the intelligibility of band-limited speech degraded by additive noise, and 
is calculated by computing a weighted average of the signal-to-noise ratio (SNR) of several frequency bands. More recently, the AI has been refined to incorporate the results of new experiments and is now known as the speech intelligibility index (SII) \cite{sii12}.

Another intelligibility metric that was developed early on is the speech transmission index (STI) \cite{houtgast1971evaluation}. %The STI can predict intelligibility for speech distorted by additive noise and reverberation. 
For this intelligibility metric, probe signals consisting of sinusoidally modulated Gaussian noise are transmitted through the communication system. The change in the modulation depth of the probe signals at the receiver is then measured and converted to an apparent SNR for each frequency band. Subsequently, the apparent SNRs are averaged similarly to the AI and SII.

Both the SII and STI have found widespread use by engineers and audiologists. However, the SII and STI have a number of limitations. First, both metrics are based on long-term statistics. This means that they do not accurately account for degradations caused by noise sources that fluctuate over time such as competing talkers and wind \cite{rhebergen2005speech}. Second, neither metric can account for distortion introduced by enhancement algorithms \cite{ludvigsen1993evaluation, loizou2011extending}. %In \cite{ludvigsen1993evaluation} it was shown that spectral subtraction algorithms \cite{boll1979suppression} can increase modulations in speech signals, which cause the STI to increase, despite the fact that spectral subtraction often reduces intelligibility. The SII also fails in this scenario because the SNR is not clearly defined after applying post-processing enhancement \cite{loizou2011extending}.

% should say 
%Second, neither metric can account for distortion introduced by enhancement algorithms \cite{ludvigsen1993evaluation, loizou2011extending}

To overcome the limitations of the SII and STI, a number of intelligibility metrics have been proposed. Examples include the coherence SII (CSII) \cite{kates2005coherence}, the extended SII (ESII) \cite{rhebergen2005speech}, the quasi-stationary STI (QSTI) \cite{schwerin2014improved}, the normalized covariance measure (NCM) \cite{Koch1992Auditory, goldsworthy2004analysis}, {the temporal fine-structure spectrum based index (TFSS)} \cite{chen2013hilbert}, the hearing-aid speech perception index (HASPI) \cite{kates2014hearing}, the Christiansen-Pedersen-Dau metric (CPD) \cite{christiansen2010prediction}, those based on the short-time objective intelligibility measure (STOI) (e.g., \cite{taal2011algorithm, jensen2016algorithm}), those based on the speech-based envelope power spectrum model (sEPSM) (e.g., \cite{jorgensen2011predicting, jorgensen2013multi, relano2016predicting}), and those based on the glimpse proportion metric (GP) (e.g., \cite{cooke2006glimpsing, barker2007modelling, tang2016glimpse}). Many of these metrics have not been extensively tested on data sets other than those used during their development. %and thus are susceptible to being over-trained. 
Additionally, the above metrics are often heuristically motivated, which suggests that they may not generalize well to new environments and enhancement strategies. %Despite this fact, it is common for engineers to use these metrics for conditions where the metrics have not been evaluated.

Recently, information theory has been proposed as a theoretically grounded approach to model speech communication. This is a natural direction to take given that the fundamental goal of speech communication is to transfer information from a talker to a listener. Information theory has been used to design state-of-the-art speech enhancement algorithms \cite{kleijn2015simple,khademi2017intelligibility} and intelligibility metrics \cite{taghia2014objective,jensen2014speech,vankuyk2017siib}. Moreover, \cite{van2017onthe} used the information bottleneck principle \cite{tishby2000information} %, which does not rely on prior knowledge of language, auditory processing, or speech production, 
to argue that the structure of speech might be adapted to the coding capability of the mammalian auditory system (see also \cite{smith2006efficient}).  

Motivated by the fact that many intrusive intelligibility metrics have been recently proposed but have not been widely evaluated, this paper presents a study on the accuracy of 12 existing monaural intrusive intelligibility metrics. To assess the accuracy of each metric, the strength of the relationship between intelligibility and the metric is measured. The intelligibility data were obtained from 11 experiments described in the literature. The data include Dutch, Danish, and English speech that was degraded by additive noise, reverberation, and competing talkers, and subjected to pre-processing enhancement and post-processing enhancement. 

{The majority of the intelligibility metrics in this paper were developed with Germanic languages in mind, however, the studies in} \cite{jin2014development, wong2007development, xia2012evaluation, chen2011predicting} {have suggested that many intelligibility metrics can obtain good performance for Mandarin, Cantonese, and Korean.}%, although sometimes the parameters of the intelligibility metrics require adjustment.}

{In addition to evaluating the accuracy of pre-existing intelligibility metrics, this paper analyzes why the top performing metrics have high performance. Specifically, the effect of decorrelating input features, the effect of the auditory model, and the effect of using different distortion measures is investigated.}

{Previous evaluations of intrusive intelligibility metrics exist. For example} \cite{ma2009objective, taal2010predicting} {evaluated the accuracy of intelligibility metrics for noise-reduced speech, and} \cite{taal2011evaluation} {evaluated the accuracy of intelligibility metrics for speech processed by ideal time-frequency segregation (ITFS). Those evaluations each considered a single type of degradation, whereas the evaluation in this paper considers data from many real-word scenarios.

Evaluations can also be found in publications that propose new intelligibility metrics, but in terms of the number of intelligibility metrics and the number of data sets, the scope of such evaluations is smaller than the present study. 
%To our knowledge, in terms of the number of listening tests and intelligibility metrics, this paper is the most comprehensive evaluation of intelligibility metrics to date. 
Two advantages of considering a broader scope are 1) it is easier to determine why some intelligibility metrics perform better than others, and 2) it is possible to investigate the ability of intelligibility metrics to generalize to new types of distortion.}

The remainder of this paper is organized as followed. Section \ref{sec:data} describes the listening test data and Section \ref{sec:metrics} {describes intelligibility metrics from the literature. Modified intelligibility metrics are proposed in Section} \ref{sec:mod_metrics}. Performance criteria are described in Section \ref{sec:evaluation} and results are presented in Section \ref{sec:results}. Finally, Section \ref{sec:conclusions} concludes the paper.

%%%%%%%%%%%%%%%%%%%%%%%%%%%%%%%%%%%%%%%%%%%%%%%%%%%%%%%%%%%%%
%%%%%%%%%%%%%%%%%%%%%%%%%%%%%%%%%
\section{\label{sec:data}Listening test data}
%To avoid overconfidence in the accuracy of intelligibility metrics, it is important that existing metrics are evaluated on some data sets other than those used during their development. Additionally, the data sets should cover a wide range of practical listening conditions. 
This paper considers the results of 11 intelligibility studies. From these studies, 13 data sets were created. In this section, each data set is described. Table \ref{table:data} summarizes the data sets, while the accompanying references provide additional details. The naming convention for the data sets includes the first author of the publication that describes the data set in full, and an abbreviation that indicates the type of degradation or processing. The order that the data sets are presented in is such that similar data sets are grouped together.

\begin{table*}
\renewcommand{\arraystretch}{1.1}
\centering
\caption{\label{table:data} Summary of listening test data sets. $m$ is the number of listeners and $n$ is the number of listening conditions.}
%\begin{tabular}{>{\raggedright\arraybackslash}p{.33\linewidth} >{\raggedright\arraybackslash}p{.15\linewidth} >{\raggedright\arraybackslash}p{.38\linewidth} r r}
\begin{tabular}{>{\raggedright}p{.15\linewidth} >{\raggedright}p{.17\linewidth} >{\raggedright}p{.35\linewidth} c c c}
\hline \hline
	Name	 				& Degradation  			& Enhancement strategy & {Bandwidth (kHz)} & $m$ & $n$\\ 
\hline
	JensenMOD \cite{jensen2016algorithm}	& Modulated noise		& None		&{10.0}		& 12	& 60	\\
	SantosREV \cite{santos2014improved}	& Noise \& reverb			& None		&{8.0}		& 10	& 17	\\ 
	KjemsAN \cite{kjems2009role} 			& Noise 					& None		&{7.7}		& 15	& 40	\\
	KjemsITFS \cite{kjems2009role}       	& Noise					& Ideal time-frequency segregation.		&{7.7}	& 15	& 168	\\
	TaalPOST \cite{taal2011algorithm}		& Noise 		    			& Minimum mean-squared error estimate of the short-time spectral amplitude.	&{8.7}		& 15	& 15	\\
	JensenPOST \cite{jensen2012spectral}	& Noise 		    			& Minimum mean-squared error estimate of the short-time spectral amplitude.	&{4.0}		& 13	& 20	\\
	HuPOST \cite{hu2007comparative}		& Noise		    			& Spectral subtractive, sub-space, statistical model based, and Wiener-type algorithms.	&{3.5}		& 40	& 72	\\ 
	HendriksPRE \cite{hendriks2015optimal}	& Noise \& reverb 	& Optimal energy redistribution.		&{8.0}	& 8	& 20	\\
	KleijnPRE \cite{kleijn2015simple}		& Noise 	    				& Optimal energy redistribution. 	&{8.0}	& 9	& 32	\\
	CookePRE \cite{cooke2013intelligibility}	& Noise \& competing talker	& Nine pre-processing enhancement algorithms.	&{8.0}	& 175	& 60	\\
	KhademiJOINT \cite{khademi2017intelligibility}	& Noise 	    		& MVDR beamformer, Wiener filter, \& optimal energy redistribution.	&{8.0}	&7	&24	\\
	DutchMRG 	   	& - 			& JensenPOST, HendriksPRE, KleijnPRE, and KhademiJOINT merged into a single data set.	&- 	&- 	&-	\\
	DantaleMRG    	& - 			& KjemsAN, KjemsITFS, and TaalPOST merged into a single data set. 				&-	&- 	&-	\\
\hline
\end{tabular}
\end{table*}

\subsection{JensenMOD}
The first data set consists of speech degraded by noise with strong temporal modulations. In \cite{jensen2016algorithm} phrases from the Dantale II corpus \cite{wagener2003design} were degraded by ten types of noise. Four of the noise types included Track 1, 4, 6, and 7 from the ICRA noise corpus \cite{dreschler2001icra}. The ICRA signals are synthetic signals %constructed by filtering Gaussian noise with bandpass filters with time-varying gain such that the synthetic signals have
with spectral and temporal properties similar to speech. Four of the noise types were constructed by multiplying speech-shaped noise (SSN) {(i.e., Gaussian noise with a long-term power-spectrum that is similar to the power spectrum of clean speech)} with $1+\sin(2\pi f t+\phi)$ where $\phi$ is uniformly distributed between $\pm \pi$, $t$ is the sample index, and $f = $ 2, 4, 8, or 16 Hz. The final two noise sources were machine-gun noise and destroyers-operation-room noise from the NOISEX corpus \cite{varga1993assessment}. Six SNRs were chosen for each noise source so that some stimuli were unintelligible and others were perfectly intelligible. In total there are 10 noise sources $\times$ 6 SNRs = 60 conditions. Stimuli were presented to 12 normal-hearing listeners. {For each word in a given sentence, the listeners were shown ten candidate words from which they were instructed to select from.} %The sampling rate was 20 kHz. 
See \cite{jensen2016algorithm} for more details.

\subsection{SantosREV}
The second data set consists of speech corrupted by noise and reverberation. In \cite{santos2014improved}, IEEE sentences \cite{rothauser1969ieee} were degraded by three types of distortion: 1) additive noise, 2) reverberation, and 3) additive noise and reverberation. For the additive noise distortion, SSN and babble noise at SNRs of $-5$, $0$, $5$, and $10$ dB were used. For the reverberant distortion, IEEE sentences were convolved with a room impulse response with T60 = $0.3$, $0.6$, $0.8$, $1$, and $1.4$ s. For the additive noise and reverberant distortion the sentences were convolved with room impulse responses with T60 = $0.3$ and $0.6$ s and mixed with SSN at SNRs of $5$ dB and $10$ dB. In total there are 8 noise + 5 reverberant + 4 noise and reverberant = 17 conditions. Stimuli were presented to ten normal-hearing listeners. {The listeners were instructed to transcribe sentences without any additional information and the proportion of correctly identified words was recorded.}
%A sampling rate of 16 kHz was used. 
See \cite{santos2014improved} for more details.

Originally, the distorted stimuli in SantosREV were offset in time from the clean stimuli. However, time-alignment is a requirement for many intrusive intelligibility metrics. For this paper, the signals in SantosREV were aligned by finding the time-offset that maximised the cross-correlation of the clean and distorted stimuli. This resulted in significantly higher performance scores than those reported in \cite{santos2014improved}.

\subsection{KjemsAN}
The third data set consists of speech degraded by additive noise. In \cite{kjems2009role} phrases from the Dantale II corpus \cite{wagener2003design} were degraded by four types of noise: SSN, cafeteria noise, noise from a bottling factory hall, and car interior noise. The stimuli were presented to 15 normal-hearing listeners. {The listeners were instructed to transcribe sentences without any additional information and the proportion of correctly identified words was recorded.} %The sampling rate was 20 kHz.
Based on the listening test results, Kjems \emph{et al.} derived psychometric curves that relate intelligibility to SNR for each noise type. 

For this paper, KjemsAN was created by adding the noise signals to the clean Dantale II sentences at ten SNRs. The SNRs were selected by sampling the psychometric curves at intervals of 10\% intelligibility from 10\% to 100\%. In total there are 4 noise types $\times$ 10 SNRs = 40 conditions. 

\subsection{KjemsITFS}
The fourth data set consists of speech subjected to ideal time-frequency segregation processing (ITFS) \cite{brungart2006isolating}. {ITFS processing aims to eliminate the energy of a speech signal at particular time-frequency locations by multiplying the short-time Fourier transform of the speech signal with a binary gain function}. Similarly to KjemsAN, the listening experiment was conducted by Kjems \emph{et al.}, used phrases from the Dantale II corpus \cite{wagener2003design}, involved 15 normal-hearing listeners, and used the same four types of noise. For each noise type, the noisy phrases were processed by two types of ITFS called an ideal binary mask and a target binary mask. Three SNRs were used ($-60$ dB, and SNRs corresponding to 20\% and 50\% intelligibility) and eight variants of each ITFS algorithm were considered. In total there are 168 conditions. See \cite{kjems2009role} for more details.

\subsection{TaalPOST}
The fifth data set consists of speech subjected to post-processing enhancement. In \cite{taal2011algorithm} phrases from the Dantale II corpus were degraded by SSN at SNRs of $8.9$, $7.7$, $6.5$, $5.2$, and $3.1$ dB. The MMSE-STSA enhancement algorithm \cite{ephraim1984speech} and an improved version \cite{erkelens2007minimum} were applied to the noisy phrases. In total there are 5 SNRs $\times$ (2 algorithms + 1 unprocessed) = 15 conditions. Stimuli were presented to 15 normal-hearing listeners. {The listeners were instructed to transcribe sentences without any additional information, and the proportion of correctly identified words was recorded.} %The sampling rate was 20 kHz. 
%See \cite{taal2011algorithm} for more details.

\subsection{JensenPOST}
The sixth data set consists of speech subjected to post-processing enhancement. In \cite{jensen2012spectral} phrases from the Dutch version of the Hagerman test \cite{houben2014development} were degraded by SSN at SNRs of $-8$, $-6$, $-4$, $-2$, and $0$ dB and processed by three enhancement algorithms. The three algorithms compute a minimum mean-squared error estimate of the clean speech by multiplying the short-time spectral amplitude of the noisy speech with a gain function. In total there are 5 SNRs $\times$ (3 algorithms + 1 unprocessed) = 20 conditions. Stimuli were presented to 13 normal-hearing listeners. {For each word in a given sentence, the listeners were shown ten candidate words from which they were instructed to select from.} %The sampling rate was 8 kHz. 
%See \cite{jensen2012spectral} for more details.

\subsection{HuPOST}
The seventh data set consists of speech subjected to post-processing enhancement. In \cite{hu2007comparative} IEEE sentences \cite{rothauser1969ieee} were filtered by a simulated telephone channel, degraded by four noise types: babble, car, street, and train, at SNRs of $0$ and $5$ dB, and processed by eight enhancement algorithms encompassing spectral subtractive, sub-space, statistical model based and Wiener-type algorithms. In total there are 4 noise types $\times$ 2 SNRs $\times$ (8 algorithms + 1 unprocessed)= 72 conditions. Stimuli were presented to 40 normal-hearing listeners where ten listeners were used for each of the four noise types. {The listeners were instructed to transcribe sentences without any additional information and the proportion of correctly identified words was recorded.} %The sampling rate was 8 kHz. 
See \cite{hu2007comparative} for more details.

\subsection{HendriksPRE}
The eighth data set consists of speech subjected to pre-processing enhancement and degraded by reverberation and noise. In \cite{hendriks2015optimal} phrases from the Dutch version of the Hagerman test \cite{houben2014development} were processed by four enhancement algorithms, convolved with a room impulse response with a T60 time of 1 s, and then degraded by SSN at SNRs of $-2$, $0$, $2$, and $4$ dB. Three of the enhancement algorithms %\cite{hendriks2015optimal, taal2013optimal, sauert2010near}
optimally redistribute the energy of the clean speech according to a distortion criterion. The fourth algorithm %\cite{hodoshima2006improving} 
uses steady-state suppression to reduce degradation caused by reverberation. In total there are 4 SNRs $\times$ (4 algorithms + 1 unprocessed) = 20 conditions. Stimuli were presented to eight normal-hearing listeners. {For each word in a given sentence, the listeners were shown ten candidate words from which they were instructed to select from.} %The sampling rate was 16 kHz.
See \cite{hendriks2015optimal} for more details.

\subsection{KleijnPRE}
The ninth data set consists of speech subjected to pre-processing enhancement and degraded by noise. In \cite{kleijn2015simple} phrases from the Dutch version of the Hagerman test \cite{houben2014development} were subjected to three pre-processing enhancement algorithms and then degraded either by SSN at SNRs of $-15, -12, -9,$ and $-6$ dB, or car noise at SNRs of $-23, -20, -17,$ and $-14$ dB. The three enhancement algorithms optimally redistribute the energy of the clean speech according to a distortion criterion. In total there are 2 noise types $\times$ 4 SNRs $\times$ (3 algorithms + 1 unprocessed) = 32 conditions. Stimuli were presented to nine normal-hearing listeners. {For each word in a given sentence, the listeners were shown ten candidate words from which they were instructed to select from.} %The sampling rate was 16 kHz. 
See \cite{kleijn2015simple} for more details.

\subsection{CookePRE}
The tenth data set consists of speech subjected to pre-processing enhancement and degraded by noise. In \cite{cooke2013intelligibility} IEEE sentences \cite{rothauser1969ieee} were processed by 19 pre-processing enhancement algorithms and degraded either by SSN at SNRs of $1$, $-4$, and $-9$ dB, or by speech from a competing talker at SNRs of $-7$, $-14$, and $-21$ dB. Stimuli were presented to 175 normal-hearing listeners. {The listeners were instructed to transcribe sentences without any additional information and the proportion of correctly identified words was recorded. Short words (e.g., ‘a’, ‘the’, ‘in’, ‘to’) were not scored.} %The sampling rate was 16 kHz. 
%See \cite{cooke2013intelligibility} for more details. 

For this paper, a subset of the data in \cite{cooke2013intelligibility} was considered because the entire data set was not available. Ten of the IEEE sentences for each condition and nine of the enhancement algorithms were used. The algorithms are referred to in \cite{cooke2013intelligibility} as AdaptDRC, F0-shift, IWFEMD, on/offset, OptimalSII, RESSYSMOD, SBM, SEO, and SSS. In total there are 2 noise sources $\times$ 3 SNRs $\times$ (9 algorithms + 1 unprocessed) = 60 conditions.

\subsection{KhademiJOINT}
The eleventh data set consists of speech that has been jointly processed by far-end and near-end enhancement algorithms. In \cite{khademi2017intelligibility}, four enhancement strategies were considered, all of which used an MVDR beamformer at the far-end. The first strategy used no near-end enhancement, the second used blind optimal energy redistribution at the near-end, the third used blind optimal energy redistribution at the near-end and an additional Wiener filter at the far-end, and the fourth used jointly optimal energy redistribution at the near-end. Three near-end SNRs ($-7.5$, 0, and 5 dB) and two far-end SNRs ($-10$ and 2.5 dB) were used. In total there are 4 enhancement strategies $\times$ 3 near-end SNRs $\times$ 2 far-end SNRs = 24 conditions. For each condition phrases from the Dutch version of the Hagerman test \cite{houben2014development} were presented to seven normal-hearing listeners. {For each word in a given sentence, the listeners were shown ten candidate words from which they were instructed to select from.} %The sampling rate was 16 kHz. 
See \cite{khademi2017intelligibility} for more details.

\subsection{DutchMRG}
The twelfth data set was created by merging JensenPOST, HendriksPRE, KleijnPRE, and KhademiJOINT. It is reasonable to merge these data sets because the associated listening tests all used phrases from the Dutch version of the Hagerman test \cite{houben2014development} and were conducted using the same procedures by the Circuits and Systems Group at Delft University of Technology. Note, that the number of subjects differed for the four experiments. DutchMRG was included in the evaluation to test if the intelligibility metrics give consistent measurements for different enhancement strategies.

\subsection{DantaleMRG}
The thirteenth data set was created by merging KjemsAN, KjemsITFS, and TaalPOST. It is reasonable to merge these data sets because the associated listening tests all used phrases from the Dantale II corpus. To prevent KjemsITFS from dominating the other data sets, 60 out of the 168 conditions from KjemsITFS were randomly selected, and all of the conditions for KjemsAN and TaalPOST were selected. Note that the listening tests were conducted by different laboratory groups. Similarly to DutchMRG, this data set was included to test if the intelligibility metrics give consistent measurements for different enhancement strategies. {JensenMOD also used the Dantale II corpus, but was not included in DantaleMRG because the listening test for JensenMOD presented listeners with ten candidate words to select from, whereas the listening tests for KjemsAN, KjemsITFS, and TaalPOST did not.}

%%%%%%%%%%%%%%%%%%%%%%%%%%%
\section{\label{sec:metrics} Pre-Existing Intelligibility Metrics}
Over the past decade a large number of intrusive intelligibility metrics have been proposed. %The metrics typically involve two stages. First, a clean acoustic speech signal and a distorted signal are converted to a representation based on an auditory model, and second, a statistic that quantifies the similarity of the clean signal and the distorted signal is calculated. 
In this section, 12 metrics from the literature, which are considered in this evaluation, are summarized. An overview of the metrics can be found in Table \ref{table:metrics}. See the accompanying references for more detailed descriptions. {Unless stated otherwise, all parameters were selected according to those recommended in the original publications.}

\subsection{Speech Intelligibility Index}
The speech intelligibility index (SII) \cite{sii12} is based on the idea that intelligibility is related to audibility. To compute the SII, a bandpass filterbank is applied to the clean speech and the noise signal, and a weighted average of the long-term SNR of each frequency band is calculated. The weights define a band-importance function (BIF) that characterizes the relative importance of each frequency band. Prior to averaging, the SNR is clipped to be between $\pm$ 15 dB and normalized to be between 0 and 1. This reflects the idea that below $-15$ dB the speech signal is inaudible and above 15 dB the intelligibility is at its maximum. The SII is known to perform well for speech degraded by stationary additive noise, but poorly for speech degraded by modulated noise sources \cite{rhebergen2005speech}. 

In this paper, the SII was only evaluated using JensenMOD, KjemsAN, and CookePRE. For the remaining data sets, either the noise signal was not available, or noise was not the main cause of distortion. The implementation of the SII was obtained from the Acoustical Society of America (\url{http://sii.to}) and used the 1/3 octave band procedure with the BIF tabulated in Table 3 of \cite{sii12}.

\subsection{High-Energy Glimpse Proportion Metric}
The glimpse proportion metric (GP) is the initial stage of the glimpsing model of speech perception \cite{cooke2006glimpsing} and has been used as an intelligibility metric in various studies (e.g., \cite{barker2007modelling, tang2016glimpse}). The GP is defined as the proportion of spectro-temporal regions where the clean speech has energy greater than the noise signal by a pre-defined threshold. The GP shares similarities with the SII in that both metrics assume that audibility is the determining factor of intelligibility. The difference is that the SII averages the long-term SNR of each frequency band, whereas the GP is the proportion of short-time frequency-local SNRs above a threshold. 

In \cite{tang2016glimpse} a variation of the GP called the high-energy GP (HEGP) was shown to be more highly correlated with intelligibility than the original GP. The main difference between the metrics is that HEGP only uses spectro-temporal regions where the noisy speech has above average energy. Similarly to the SII, HEGP can only quantify distortion caused by additive noise signals. For this reason, HEGP was evaluated using KjemsAN, JensenMOD, and CookePRE only. 

The implementation of HEGP used in this paper was obtained from its developers. Note that CookePRE is a subset of a data set that was used during the development of HEGP.

%place table 2 here
\begin{table}[]
\renewcommand{\arraystretch}{1.3}
\caption{\label{table:metrics} Pre-existing intelligibility metrics considered in this study.}
\begin{tabular}{l >{\raggedright\arraybackslash}p{.7\linewidth}}
\hline \hline
	Abbreviation 	& Description  \\ \hline
       	SII		       	& The speech intelligibility index \cite{sii12}.  	    \\
       	HEGP		& The high-energy glimpse proportion metric \cite{tang2016glimpse}. 	    \\
       	CSII-MID	& The mid-level coherence SII \cite{kates2005coherence}.  	    \\
       	HASPI		& The hearing-aid speech perception index \cite{kates2014hearing}. 	    \\ 
	NCM-BIF	& The normalized covariance measure with signal-dependent band-importance functions \cite{ma2009objective}.  	    \\
	QSTI	       	& The quasi-stationary speech transmission index \cite{schwerin2014improved}. 	    \\
	STOI		& The short-time objective intelligibility measure \cite{taal2011algorithm}. 	    \\
	ESTOI		& The extended STOI measure \cite{jensen2016algorithm}. 	    \\ 
	MIKNN	       	& The k-nearest neighbour mutual information intelligibility measure \cite{taghia2014objective}. 	    \\
	SIMI		& Speech intelligibility prediction based on a mutual information lower bound \cite{jensen2014speech}. 	    \\
	SIIB			& Speech intelligibility in bits \cite{vankuyk2017siib}.		    \\
	$\text{sEPSM}^\text{corr}$ 	& The speech-based envelope power spectrum model with short-time correlation \cite{relano2016predicting}.    \\
	\hline \hline
\end{tabular}
\end{table}

\subsection{Coherence Speech Intelligibility Index}
The coherence speech intelligibility index (CSII) \cite{kates2005coherence} is based on the SII, but replaces the SNR of each frequency band with a signal-to-distortion ratio (SDR). The SDR is estimated from the coherence function \cite{carter1973estimation} of the clean and distorted speech signal. For the case of speech degraded by additive noise, the SDR and SNR are equivalent, making the CSII a generalization of the SII that can be applied to a wider range of distortions. In \cite{kates2005coherence} it was found that the performance of the CSII could be improved by calculating the CSII separately for low, mid, and high-energy speech segments. %, and then taking a weighted combination of the three CSII values.

The implementation of the CSII used in this paper was obtained from \cite{loizou2013speech} and is described in \cite{ma2009objective}, where it is referred to as $\mathrm{CSII_{mid}}$. Note that the implementation in \cite{loizou2013speech} differs to that originally proposed in \cite{kates2005coherence} in that \cite{loizou2013speech} averages the CSII over short-time segments. For this paper, the implementation in \cite{loizou2013speech} was modified to make it more similar to that originally proposed (i.e., it does not use short-time segments) because we found that the original method had higher overall performance. In this paper the algorithm is referred to as CSII-MID.

\subsection{Hearing-Aid Speech Perception Index}
The hearing-aid speech perception index (HASPI) \cite{kates2014hearing} is based on an elaborate auditory model where the shape and bandwidth of the cochlear filters depend on the speech signal intensity and the outer hair-cell damage of the listener. Dynamic range compression is applied to the output of each cochlear filter in accordance with physiological measurements of compression in the cochlea and psychophysical estimates of compression in the human ear. Additionally, a time-alignment stage is included. The auditory model has two outputs: a sequence of short-time log-spectra, and a basilar membrane vibration signal for each frequency band.

From the outputs of the auditory model the cepstral correlation and auditory coherence are computed. To compute cepstral correlation, the log-spectra are converted to an approximation of Mel-frequency cepstral coefficients \cite{davis1980comparison} by taking the inner product between the log-spectra and a set of cosine functions. Pearson's correlation coefficient between the cepstra of the clean and distorted speech is then computed for each cepstral dimension and the resulting coefficients are averaged.

The auditory coherence is computed by splitting the basilar membrane vibration signals into three sets that contain low, mid, and high-energy segments. For each set and each frequency band, short-time correlation coefficients between the clean vibration signals and the distorted vibration signals are computed and then averaged over the time dimension and the frequency dimension. This results in three auditory coherence terms corresponding to low, mid, and high energy segments.

HASPI is computed as a linear combination of the cepstral correlation and the three auditory coherence terms. The relative importance of each term depends on the type of distortion and thus is fitted to the intelligibility data. In this paper the weights of the cepstral correlation and auditory coherence terms were computed for each data set such that the mean squared error between the predicted and measured intelligibility scores was minimized. However, it was found that similar performance could be obtained simply by summing the cepstral correlation and high-energy auditory coherence. 
The implementation of HASPI used in this paper was obtained from its developers.

\subsection{Normalized Covariance Measure}
The normalized covariance measure (NCM) \cite{Koch1992Auditory, goldsworthy2004analysis} is a variant of the STI that uses clean speech as the probe signal. To compute the NCM, a band-pass filterbank is applied to the clean and distorted speech signals, and the temporal envelope of the output of each filter is extracted. Subsequently, the normalized covariance (i.e., Pearson's correlation coefficient) between the clean and distorted envelopes is calculated and converted to an apparent SNR for each frequency band. Similarly to the SII, the apparent SNR is clipped before a weighted average over the frequency bands is computed. 

In \cite{ma2009objective} it was found that the NCM is strongly correlated with intelligibility for speech subjected to post-processing enhancement. The correlation was particularly strong when new signal dependent BIFs were used. The implementation of the NCM used in this paper was obtained from \cite{loizou2013speech} and is described in \cite{ma2009objective} where it is referred to as NCM $W^{(1)}_i, p=1.5$. In this paper the algorithm is referred to as NCM-BIF. Note that HuPOST was used during the development of NCM-BIF.

\subsection{Quasi-Stationary Speech Transmission Index}
The quasi-stationary speech transmission index (QSTI) was proposed in \cite{schwerin2014improved}. The QSTI is a variation of the STI that uses clean speech as the probe signal and averages the score over short-time segments. In \cite{schwerin2014improved} the QSTI was reported to be more strongly correlated with intelligibility than the traditional STI. 

The implementation of the QSTI used in this paper was obtained from its developers webpage. Note that HuPOST, TaalPOST, and KjemsITFS were used during the development of QSTI.

\subsection{Short-Time Objective Intelligibility Measure}
The short-time objective intelligibility measure (STOI) was proposed in \cite{taal2011algorithm} as an algorithm for predicting the intelligibility of time-frequency weighted noisy speech. To compute STOI, a simple model of the human auditory system is used to extract temporal envelopes of the clean speech and the distorted speech for various frequency bands. The temporal envelopes are segmented into short-time frames with a duration of 386 ms and a clipping procedure is used to ensure that the SDR of each frame is greater than $-15$ dB. STOI is calculated by computing Pearson's correlation coefficient between the clean and distorted envelopes for each short-time frame and each frequency band and then taking the mean.

%STOI performs well for speech distorted by stationary noise \cite{jensen2014speech}, post-processing enhancement \cite{taal2011algorithm}, and mobile phones \cite{jorgensen2015speech}, but poorly for speech distorted by modulated noise sources \cite{jensen2016algorithm}. 
The implementation of STOI used in this paper was obtained from its developer's webpage. Note that TaalPOST and KjemsITFS were used during the development of STOI.

\subsection{Extended Short-Time Objective Intelligibility Measure}
The extended short-time objective intelligibility measure (ESTOI) was proposed in \cite{jensen2016algorithm} to address the finding that STOI performs poorly for modulated noise sources (e.g., Gaussian noise that is amplitude modulated by a sinusoid). Rather than computing the correlation of the clean and distorted envelopes for short-time segments, ESTOI computes the correlation between clean and distorted spectra so that `glimpses of clean speech' can be detected. Additionally, the clipping procedure in STOI was removed to make the new model more mathematically tractable. %\cite{jensen2016algorithm} found that ESTOI has higher performance than STOI for speech distorted by modulated noise, and similar performance to STOI for other conditions. 

The implementation of ESTOI used in this paper was obtained from its developer's webpage. Note that JensenPOST, JensenMOD, KjemsITFS, and a data set similar to KjemsAN were used during the development of ESTOI. 

\subsection{K-Nearest Neighbour Mutual Information Intelligibility Measure}
The k-nearest neighbour (KNN) mutual information intelligibility measure (MIKNN) was proposed in \cite{taghia2014objective} while investigating the use of information theoretical techniques for intelligibility prediction. MIKNN uses the same representation of speech as STOI, however, rather than using the short-time correlation coefficient to quantify distortion, MIKNN estimates the mutual information between the clean and distorted temporal envelopes using a non-parametric estimator based on k-nearest neighbours \cite{kraskov2004estimating}. One advantage of mutual information is that unlike Pearson's correlation coefficient, mutual information can account for non-linear dependencies.

The implementation of MIKNN used in this paper was obtained from its developer's webpage. Note that TaalPOST and KjemsITFS were used during the development of MIKNN.

\subsection{Speech Intelligibility Prediction Based on Mutual Information}
Similarly to MIKNN, the speech intelligibility prediction based on mutual information measure (SIMI) \cite{jensen2014speech} is based on the hypothesis that intelligibility is related to the mutual information between the clean and distorted temporal envelopes. In contrast to MIKNN, SIMI estimates a lower bound on the mutual information by assuming a parametric statistical model. Another important difference between SIMI and MIKNN is that SIMI operates on short-time segments of 250 ms, whereas MIKNN uses whole utterances. In \cite{jensen2014speech} SIMI was used to justify some of the heuristic design decisions of STOI.

The implementation of SIMI used in this paper was obtained from its developer's webpage. Note that JensenPOST, KjemsITFS, and a data set similar to KjemsAN were used during the development of SIMI.

\subsection{Speech Intelligibility in Bits}
Speech intelligibility in bits (SIIB) is an information theoretic intelligibility metric that was recently proposed in \cite{vankuyk2017siib}. Similar to MIKNN, a non-parametric mutual information estimator \cite{kraskov2004estimating} is used to estimate the information shared between a clean and distorted speech signal.

There are three main differences between SIIB and MIKNN. First, SIIB uses the Karhunen-Lo\`eve transform (KLT) \cite{karhunen1947lineare} {to reduce statistical dependencies between spectro-temporal regions, and thus reduces overestimation of the information rate.} 

Second, SIIB accounts for `production noise', which incorporates differences in pronunciation between talkers. Importantly, production noise causes the information rate of the communication channel to saturate \cite{kleijn2015simple}. 

Third, SIIB uses an auditory model that more accurately accounts for the frequency masking \cite{wegel1924auditory} and temporal masking \cite{oxenham2001forward} of the human auditory system. To account for frequency masking, the temporal envelopes are extracted using an equivalent rectangular bandwidth  (ERB) gammatone filterbank \cite{slaney1993efficient}. To account for temporal masking, the forward masking function suggested in \cite{rhebergen2006extended} is used. Additionally, logarithmic compression is applied to the envelopes.

The end result of SIIB is an estimate of the information shared between a talker and a listener in bits per second. Note that all of the data sets considered in this paper were used during the development of SIIB.

\subsection{Speech-Based Envelope Power Spectrum Model with Short-Time Correlation}
The speech-based envelope power spectrum model forms the basis of three intelligibility metrics: sEPSM \cite{jorgensen2011predicting}, mr-sEPSM \cite{jorgensen2013multi}, and $\text{sEPSM}^\text{corr}$ \cite{relano2016predicting}. All of the sEPSM metrics use the Hilbert transform and a gammatone filterbank to extract temporal envelopes for different frequency bands. A second bandpass filterbank called a modulation filterbank is then applied to each envelope signal. This results in a multi-dimensional representation that includes a time, frequency, and modulation dimension. Within this multi-dimensional domain, sEPSM and mr-sEPSM quantify distortion using a SNR metric, whereas $\text{sEPSM}^\text{corr}$ quantifies distortion using short-time correlation coefficients similarly to STOI. In this paper only the most recent metric is considered: $\text{sEPSM}^\text{corr}$. 

Note that the output of $\text{sEPSM}^\text{corr}$ increases as the duration of the stimulus increases. This is a consequence of the `multiple looks' strategy that $\text{sEPSM}^\text{corr}$ uses to integrate information over the time dimension. For this reason, when comparing results from multiple data sets (i.e., for the merged data sets), it is important that the duration of the stimuli is held constant. In this paper, when evaluating $\text{sEPSM}^\text{corr}$, all stimuli were truncated to have a duration of 20 seconds. 

The implementation of $\text{sEPSM}^\text{corr}$ used in this paper was obtained from its developers. Note that KjemsITFS was used during the development of $\text{sEPSM}^\text{corr}$.

\section{\label{sec:mod_metrics}{Modified Intelligibility Metrics}}
%Note that the intelligibility metrics in the previous section typically involve three stages: 1) a clean acoustic speech signal and a distorted signal are converted to a representation based on an auditory model, 2) a statistic that quantifies the similarity of the signals is calculated for time and/or frequency local regions, and 3) the similarity of the regions are summed to give a final scalar value. 
One of the goals of this paper is to investigate why some intelligibility metrics have higher performance than others. In this section we modify existing intelligibility metrics so that effective strategies can be identified.

\subsection{{Investigating the effect of decorrelating input features}}
The majority of the intelligibility metrics in the previous section quantify distortion by comparing time and/or frequency local features. SIIB and HASPI are exceptions to this. SIIB decorrelates log-spectra over the time and frequency dimension using the KLT, and HASPI decorrelates log-spectra over the frequency dimension using a cosine expansion similar to the type-1 discrete cosine transform (DCT) \cite{rao2014discrete}. Recall that for stationary signals the DCT asymptotically approximates the KLT. 

To investigate the effect of decorrelating input features, SIIB and STOI were modified to produce two intelligibility metrics denoted $\text{SIIB}^\text{noKLT}$ and $\text{STOI}^\text{KLT}$. To compute $\text{SIIB}^\text{noKLT}$, the implementation of SIIB described in \cite{vankuyk2017siib} was used, but the KLT was not applied. To compute $\text{STOI}^\text{KLT}$ three changes are made to the original STOI implementation \cite{taal2011algorithm}:
\begin{enumerate}
\item Instead of using temporal envelopes to represent speech signals, log-temporal envelopes are used. To prevent singularities, a small amount of uniformly distributed noise is added to the envelopes before applying the logarithm.
\item The KLT is used to decorrelate the log-temporal envelopes over the frequency dimension. To do so, the eigenvectors of the covariance matrix of the clean log-temporal envelopes are computed.
\item Short-time correlation coefficients for the eigenchannels are computed and then averaged to produce a final value. The short-time segmentation approach in \cite{taal2011algorithm} is used, but the clipping procedure is not.
\end{enumerate} 
By comparing the performance of STOI with $\text{STOI}^\text{KLT}$, and SIIB with $\text{SIIB}^\text{noKLT}$ the effect of decorrelating input features can be investigated.

\subsection{{Investigating the effect of the auditory model}}
The auditory model that is used to extract features could have a significant impact on performance. To investigate this effect, the auditory model used for $\text{STOI}^\text{KLT}$ (i.e., STOIs auditory model) was replaced with the auditory model used by SIIB. The differences between the auditory models are: 1) SIIB uses an ERB gammatone filterbank, whereas STOI uses a 1/3 octave band rectangular filterbank, 2) SIIB considers frequencies up to 8 kHz, whereas STOI considers frequencies up to 5 kHz, and 3) SIIB includes a forward temporal masking function, whereas STOI does not. The resulting intelligibility metric is denoted $\text{STOI}^\text{KLT}_\text{gamma}$.

\subsection{{Investigating the effect of mutual information estimation}}
The majority of the intelligibility metrics in the previous section rely on the correlation coefficient to quantify distortion. On the other hand, SIIB and MIKNN use a non-parametric mutual information estimator. %Unlike the correlation coefficient, mutual information can quantify non-linear dependencies between random variables. 
Recall that if the clean and degraded signals are jointly Gaussian, then the mutual information is a function of the correlation coefficient only. In \cite{jensen2014speech} this observation was used to justify the use of the correlation coefficient. However, a direct comparison between the performance obtained using a non-parametric mutual information estimator and the performance obtained using the capacity of a Gaussian channel has not been made. 

To investigate the effect of mutual information estimation, SIIB was modified to produce a simpler metric called $\text{SIIB}^\text{Gauss}$. The original SIIB algorithm \cite{vankuyk2017siib} quantifies distortion using a KNN mutual information estimator, whereas $\text{SIIB}^\text{Gauss}$ uses the information capacity of a Gaussian channel. Concretely, \begin{equation} \text{SIIB}^\text{Gauss} = -\frac{F}{2K}\sum_j \log_2(1-r^2\rho_j^2),\end{equation} where $F$ is the frame rate, $K=15$ is the number of stacked log-spectra, $r=0.75$ is the production noise correlation coefficient, $j$ is the eigenchannel index, and $\rho_j$ is the correlation coefficient between the $j$th clean eigenchannel and the $j$th distorted eigenchannel. The values for $F$, $K$ and $r$ are the same as those in \cite{vankuyk2017siib}.

%%%%%%%%%%%%%%%%%%%%%%%%%%%%%%%%%
\section{\label{sec:evaluation}Performance Criteria}
The key requirement of an intelligibility metric is that it has a strong monotonic increasing relationship with intelligibility. %Then, if a change in a communication system results in an increase in the intelligibility metric, an increase in intelligibility can also be expected. 
This paper uses two performance criteria to quantify the strength of the relationship: Kendall's tau coefficient, $\tau$, and Pearson's correlation coefficient, $\rho$. Both performance criteria are discussed below.

In the following, $p_c$ is the intelligibility in terms of percentage of words correctly identified for condition $c$ in a particular data set and $d(x_c,y_c)$ is the corresponding score computed by an intelligibility metric. The clean signal $x_c$ is formed by concatenating all available clean sentences for condition $c$ and likewise for the distorted signal $y_c$.

\subsection{Kendall's Tau Coefficient}
Kendall's tau coefficient \cite{kendall1938new}, $\tau$, measures the ordinal association between two quantities and %Let $i$ and $j$ be two conditions in a data set where $i\neq j$. The pair formed by $(w_i,d(x_i,y_i))$ and $(w_{j},d(x_j,y_j))$ is concordant if both $w_i>w_{j}$ and $d(x_i,y_i)>d(x_j,y_j)$, or if both $w_i<w_{j}$ and $d(x_i,y_i)<d(x_j,y_j)$. The pair is disconcordant if $w_i>w_{j}$ and $d(x_i,y_i)<d(x_j,y_j)$ or if $w_i<w_{j}$ and $d(x_i,y_i)>d(x_j,y_j)$. Kendall's tau coefficient is given by
%\begin{equation}
%	\tau = \dfrac{n_C-n_D}{n(n-1)/2},
%\end{equation} 
%where $n_C$ is the number of concordant pairs, $n_D$ is the number of disconcordant pairs, and $n$ is the number of conditions in the data set. Kendall's tau coefficient 
ranges between $-$1 and 1. If  $\tau=-1$ then $p_c$ and $d(x_c,y_c)$ have a monotonic decreasing relationship, if $\tau=1$ they have a monotonic increasing relationship, and if they are statistically independent then $\tau=0$.

\subsection{Pearson's Correlation Coefficient}
Pearson's correlation coefficient, $\rho$, is defined as the normalized covariance between two quantities. To use $\rho$ effectively, the relationship between the quantities must be linear. For this reason, a monotonic function $f$ is applied to $d(x_c,y_c)$ to linearize the relationship before computing $\rho$. The function $f$ can be thought of as a mapping from the metric to predicted intelligibility scores, but more generally it is simply a tool for quantifying the strength of the relationship between $d(x_c,y_c)$ and $p_c$.

{In the literature $f$ is commonly assumed to be a logistic function, e.g., } \cite{gordon1995comparing,kates2005coherence,taal2011algorithm}:
\begin{equation}
	f(d(x_c,y_c)) = \dfrac{100}{1+e^{a(d(x_c,y_c)-b)}},
	\label{eq:sigmoid_old}
\end{equation}
{where $b$ is the midpoint and $a$ is the slope at the midpoint. These parameters are fitted to the data to minimize the mean squared error between $p_c$ and $f(d(x_c,y_c))$.} %However, \eqref{eq:sigmoid_old} {does not always fit the data well, thus in this paper a more flexible function is used} \cite{fletcher1950perception,studebaker1991frequency}:
%\begin{equation}
%	f(d(x_c,y_c)) = 100(1-e^{-ad(x_c,y_c)})^b,
%	\label{eq:sigmoid}
%\end{equation}
%where $a,b>0$. Figure \ref{fig:mapping} shows several examples of \eqref{eq:sigmoid_old} and \eqref{eq:sigmoid}. The advantage of \eqref{eq:sigmoid} is that the curve does not have to be symmetric about it's midpoint. 

In the literature $\rho$ is sometimes also computed without applying a mapping function. However, we believe that such a measure is misleading because without $f$, a metric with a strong non-linear relationship between $p_c$ and $d(x_c,y_c)$ will have a small value for $\rho$, but could also have a monotonic increasing relationship with intelligibility. 

%\begin{figure}[t!]
%\centering
%\includegraphics[scale=1]{figure1}
%\vspace*{-0.2cm}
%\caption{Left: examples of the mapping in \eqref{eq:sigmoid_old}. Right: examples of the mapping in \eqref{eq:sigmoid}.}
%\label{fig:mapping}
%\vspace{-0.2cm}
%\end{figure}

Note that $p_c$ depends on the experimental procedures used to measure intelligibility, but that $d(x_c,y_c)$ does not. For example, the intelligibility of a given stimulus can be increased by changing an open listening test to a closed listening test\footnote{{In a closed listening test, subjects are given a list of possible speech sounds, e.g., phones or words, and are asked to identify the sounds that they heard. In an open listening test, no list is provided, which makes the test more difficult.}}. It follows that the relationships between intelligibility and intelligibility metrics also depend on experimental procedures. For this reason, $f$ is fit individually to each data set. %Pearson's correlation coefficient is calculated according to
%\begin{equation}
%	\rho = \dfrac{\sum_c (p_c-\bar{w}_c)(f(d(x_c,y_c))-\bar{f}(d(x_c,y_c)))}{\sqrt{\sum_c (p_c-\bar{w}_c)^2 \sum_c (f(d(x_c,y_c))-\bar{f}(d(x_c,y_c))})^2},
%\end{equation}
%where the overbar is used to denote the mean over all conditions in the data set. 
%Finally, because the relationship between an intelligibility metric and intelligibility should be monotonically increasing, 
Finally, negative values of $\rho$ and $\tau$ are set to zero.

\section{\label{sec:results}Results}
\begin{figure*}[h!]
\centering
\includegraphics[width=\textwidth]{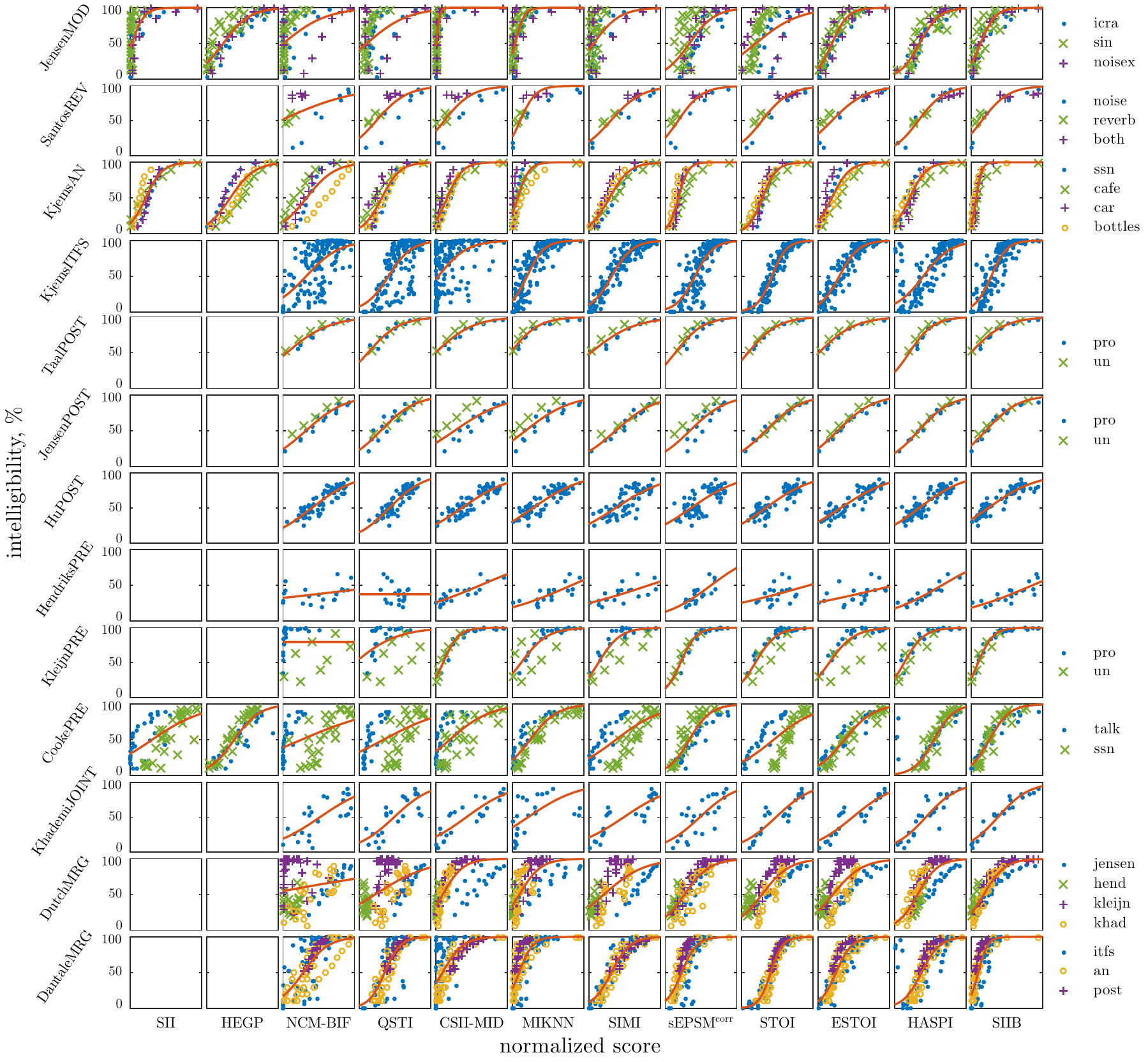}
\vspace*{-0.2cm}
\caption{Scatter plots for all data sets and pre-existing intelligibility metrics. {The vertical axis is the 'ground-truth' intelligibility in terms of the percentage of words correctly identified during listening tests}, and the horizontal axis is the score computed by an intelligibility metric. The horizontal axis of each plot has been normalized to be between 0 and 1. {Each data point corresponds to a processing condition}. The mapping function in \eqref{eq:sigmoid_old} is also shown.}
\label{fig:scatter}
\end{figure*}

Scatter plots for all data sets described in Section II and all pre-existing intelligibility metrics described in Section III are displayed in Figure \ref{fig:scatter}. Each row of plots corresponds to a data set and each column of plots corresponds to an intelligibility metric. {The vertical axis of each scatter plot is the 'ground-truth' intelligibility in terms of the percentage of words correctly identified during listening tests}, and the horizontal axis is the score computed by an intelligibility metric. To facilitate an easy visual comparison, the horizontal axis of each scatter plot is normalized to be between 0 and 1. Each point on a scatter plot corresponds to a condition in the respective data set. The function in \eqref{eq:sigmoid_old} that was used to linearize the relationship between the intelligibility scores and the metric for each data set is also shown. For an ideal intelligibility metric, all points would fall exactly on top of the fitted curve. 

\begin{table*}[t!]
\centering
\renewcommand{\arraystretch}{1.2}
\caption{\label{table:tau} Performance in terms of Kendall's tau coefficient, $\tau$, for all data sets and intelligibility metrics. {The intelligibility metrics are listed in order of mean performance and are grouped by pre-existing metrics (left) and modified metrics (right).}}
\begin{tabular}{l dddddddddddd|dddd}
\hline\hline
& \multicolumn{1}{p{0.55cm}}{\rotatebox[origin=c]{50}{\text{SII}}} 
& \multicolumn{1}{p{0.55cm}}{\rotatebox[origin=c]{50}{\text{HEGP}}} 
& \multicolumn{1}{p{0.55cm}}{\rotatebox[origin=c]{50}{\text{NCM-BIF}}} 
& \multicolumn{1}{p{0.55cm}}{\rotatebox[origin=c]{50}{\text{QSTI}}} 
& \multicolumn{1}{p{0.55cm}}{\rotatebox[origin=c]{50}{\text{CSII-MID}}} 
& \multicolumn{1}{p{0.55cm}}{\rotatebox[origin=c]{50}{\text{MIKNN}}} 
& \multicolumn{1}{p{0.55cm}}{\rotatebox[origin=c]{50}{\text{SIMI}}} 
& \multicolumn{1}{p{0.55cm}}{\rotatebox[origin=c]{50}{$\text{sEPSM}^\text{corr}$}}
& \multicolumn{1}{p{0.55cm}}{\rotatebox[origin=c]{50}{\text{STOI}}} 
& \multicolumn{1}{p{0.55cm}}{\rotatebox[origin=c]{50}{\text{ESTOI}}} 
& \multicolumn{1}{p{0.55cm}}{\rotatebox[origin=c]{50}{\text{HASPI}}} 
& \multicolumn{1}{p{0.55cm}}{\rotatebox[origin=c]{50}{\text{SIIB}}} 
& \multicolumn{1}{p{0.55cm}}{\rotatebox[origin=c]{50}{$\text{SIIB}^\text{noKLT}$}} 
& \multicolumn{1}{p{0.55cm}}{\rotatebox[origin=c]{50}{$\text{STOI}^\text{KLT}$}} 
& \multicolumn{1}{p{0.55cm}}{\rotatebox[origin=c]{50}{$\text{STOI}^\text{KLT}_\text{gamma}$}} 
& \multicolumn{1}{p{0.55cm}}{\rotatebox[origin=c]{50}{$\text{SIIB}^\text{Gauss}$}} 
\\        
\hline 
JensenMOD 	& 0.52 	& 0.71 	& 0.41 	& 0.34 	& 0.57 	& 0.55 	& 0.34 	& 0.51 	& 0.38 	& 0.75^*& 0.75 	& 0.74^* 		& 0.59	& 0.72	& 0.71	& 0.74\\
SantosREV 	& - 	& - 	& 0.38 	& 0.61 	& 0.57 	& 0.70 	& 0.72 	& 0.72 	& 0.82 	& 0.79 	& 0.85 	& 0.82^* 		& 0.82	& 0.79 	& 0.79 	& 0.80\\
KjemsAN 	& 0.76 	& 0.75 	& 0.65 	& 0.78 	& 0.80 	& 0.65 	& 0.81^*& 0.74 	& 0.81 	& 0.74^*& 0.79 	& 0.82^* 		& 0.74	& 0.74 	& 0.76 	& 0.84\\
KjemsITFS 	& - 	& - 	& 0.48 	& 0.51^*& 0.41 	& 0.71^*& 0.80^*& 0.70^*& 0.82^*& 0.81^*& 0.66 	& 0.73^* 		& 0.69	& 0.73 	& 0.74 	& 0.73\\
TaalPOST 	& - 	& - 	& 0.85 	& 0.87^*& 0.81 	& 0.83^*& 0.81 	& 0.79 	& 0.92^*& 0.96 	& 0.83 	& 0.87^* 		& 0.87	& 0.79 	& 0.79 	& 0.87\\
JensenPOST 	& - 	& - 	& 0.81 	& 0.80 	& 0.60 	& 0.68 	& 0.92^*& 0.66 	& 0.89 	& 0.83^*& 0.95 	& 0.92^* 		& 0.65	& 0.82 	& 0.82 	& 0.94\\
HuPOST 		& - 	& - 	& 0.67^*& 0.68^*& 0.63 	& 0.64 	& 0.55 	& 0.44 	& 0.59 	& 0.69 	& 0.61 	& 0.74^* 		& 0.39	& 0.72 	& 0.70 	& 0.73\\ 
HendriksPRE & - 	& - 	& 0.30 	& 0.00 	& 0.69 	& 0.56 	& 0.52 	& 0.59 	& 0.26 	& 0.43 	& 0.78 	& 0.66^* 		& 0.72	& 0.53 	& 0.62 	& 0.60\\ 
KleijnPRE 	& - 	& - 	& 0.13 	& 0.20 	& 0.86 	& 0.71 	& 0.57 	& 0.88 	& 0.70 	& 0.58 	& 0.79 	& 0.86^* 		& 0.77	& 0.78 	& 0.88 	& 0.86\\
CookePRE 	& 0.44 	& 0.72^*& 0.38  & 0.38 	& 0.46 	& 0.72 	& 0.52 	& 0.71 	& 0.56 	& 0.77 	& 0.75 	& 0.76^* 		& 0.71	& 0.87 	& 0.84 	& 0.77\\ 
KhademiJOINT& - 	& - 	& 0.50 	& 0.51 	& 0.71 	& 0.53 	& 0.74 	& 0.60 	& 0.79 	& 0.80 	& 0.77 	& 0.89^* 		& 0.90	& 0.82 	& 0.87 	& 0.90\\ 
DutchMRG 	& - 	& - 	& 0.13 	& 0.29 	& 0.57 	& 0.54 	& 0.44 	& 0.68 	& 0.59 	& 0.46 	& 0.64 	& 0.75^* 		& 0.58	& 0.54 	& 0.67 	& 0.74\\ 
DantaleMRG 	& - 	& - 	& 0.54 	& 0.64 	& 0.53 	& 0.61 	& 0.80 	& 0.66 	& 0.83 	& 0.75 	& 0.67 	& 0.68^* 		& 0.58	& 0.70 	& 0.73 	& 0.71\\
&&&&&&&&&&&&
\\[-2mm]
Mean & 0.57 & 0.73 & 0.48 & 0.51 & 0.63 & 0.65 & 0.66 & 0.67 & 0.69 & 0.72 & 0.76 & 0.79 							& 0.69 & 0.73 & 0.76 & 0.79\\
$\mathrm{CI_{low}}$ & 0.50 & 0.68 & 0.43 & 0.46 & 0.59 & 0.61 & 0.61 & 0.63 & 0.65 & 0.68 & 0.72 & 0.75				& 0.66 & 0.70 & 0.73 & 0.76\\
$\mathrm{CI_{high}}$ & 0.64 & 0.77 & 0.52 & 0.55 & 0.67 & 0.69 & 0.69 & 0.70 & 0.73 & 0.75 & 0.78 & 0.81			& 0.72 & 0.76 & 0.79 & 0.81\\
\hline\hline
\end{tabular} 
\end{table*} 

\begin{table*}[t!]
\centering
\renewcommand{\arraystretch}{1.2}
\caption{\label{table:rho} Performance in terms of Pearson's correlation coefficient, $\rho$, for all data sets and intelligibility metrics. {The intelligibility metrics are listed in order of mean performance and are grouped by pre-existing metrics (left) and modified metrics (right).}}
\begin{tabular}{l dddddddddddd | dddd}  
\hline\hline
& \multicolumn{1}{p{0.55cm}}{\rotatebox[origin=c]{50}{\text{SII}}} 
& \multicolumn{1}{p{0.55cm}}{\rotatebox[origin=c]{50}{\text{HEGP}}} 
& \multicolumn{1}{p{0.55cm}}{\rotatebox[origin=c]{50}{\text{NCM-BIF}}} 
& \multicolumn{1}{p{0.55cm}}{\rotatebox[origin=c]{50}{\text{QSTI}}} 
& \multicolumn{1}{p{0.55cm}}{\rotatebox[origin=c]{50}{\text{CSII-MID}}} 
& \multicolumn{1}{p{0.55cm}}{\rotatebox[origin=c]{50}{\text{MIKNN}}} 
& \multicolumn{1}{p{0.55cm}}{\rotatebox[origin=c]{50}{\text{SIMI}}} 
& \multicolumn{1}{p{0.55cm}}{\rotatebox[origin=c]{50}{$\text{sEPSM}^\text{corr}$}}
& \multicolumn{1}{p{0.55cm}}{\rotatebox[origin=c]{50}{\text{STOI}}} 
& \multicolumn{1}{p{0.55cm}}{\rotatebox[origin=c]{50}{\text{ESTOI}}} 
& \multicolumn{1}{p{0.55cm}}{\rotatebox[origin=c]{50}{\text{HASPI}}} 
& \multicolumn{1}{p{0.55cm}}{\rotatebox[origin=c]{50}{\text{SIIB}}} 
& \multicolumn{1}{p{0.55cm}}{\rotatebox[origin=c]{50}{$\text{SIIB}^\text{noKLT}$}} 
& \multicolumn{1}{p{0.55cm}}{\rotatebox[origin=c]{50}{$\text{STOI}^\text{KLT}$}} 
& \multicolumn{1}{p{0.55cm}}{\rotatebox[origin=c]{50}{$\text{STOI}^\text{KLT}_\text{gamma}$}} 
& \multicolumn{1}{p{0.55cm}}{\rotatebox[origin=c]{50}{$\text{SIIB}^\text{Gauss}$}}
\\           
\hline
JensenMOD 	& 0.65 	& 0.88 	& 0.45 	& 0.43 	& 0.65 	& 0.72 	& 0.51 	& 0.68 	& 0.47 	& 0.92^*& 0.92 	& 0.89^* 		& 0.78 & 0.90 & 0.88 & 0.89 \\
SantosREV 	& - 	& - 	& 0.46 	& 0.76 	& 0.72 	& 0.90 	& 0.94 	& 0.87 	& 0.94 	& 0.91 	& 0.97 	& 0.93^* 		& 0.98 & 0.93 & 0.95 & 0.93 \\  
KjemsAN 	& 0.89 	& 0.89 	& 0.80 	& 0.90 	& 0.92 	& 0.78 	& 0.93^*& 0.87 	& 0.93 	& 0.87^*& 0.93 	& 0.94^* 		& 0.88 & 0.88 & 0.89 & 0.94 \\
KjemsITFS 	& - 	& - 	& 0.67 	& 0.72^*& 0.49 	& 0.88^*& 0.95^*& 0.84^*& 0.96^*& 0.95^*& 0.78 	& 0.89^* 		& 0.83 & 0.89 & 0.91 & 0.89 \\ 
TaalPOST 	& - 	& - 	& 0.95 	& 0.95^*& 0.93 	& 0.95^*& 0.92 	& 0.90 	& 0.98^*& 0.97 	& 0.95 	& 0.96^* 		& 0.96 & 0.92 & 0.92 & 0.96 \\ 
JensenPOST 	& - 	& - 	& 0.95 	& 0.93 	& 0.78 	& 0.86 	& 0.97^*& 0.80 	& 0.99 	& 0.97^*& 0.99 	& 0.98^* 		& 0.77 & 0.95 & 0.96 & 0.98 \\
HuPOST 		& - 	& - 	& 0.89^*& 0.89^*& 0.89 	& 0.88 	& 0.77 	& 0.73 	& 0.87 	& 0.90 	& 0.88 	& 0.92^* 		& 0.65 & 0.91 & 0.92 & 0.92 \\ 
HendriksPRE & - 	& - 	& 0.29 	& 0.00 	& 0.86 	& 0.76 	& 0.66 	& 0.78 	& 0.35 	& 0.47 	& 0.92 	& 0.82^* 		& 0.91 & 0.65 & 0.77 & 0.73 \\
KleijnPRE 	& - 	& - 	& 0.00 	& 0.34 	& 0.98 	& 0.82 	& 0.87 	& 0.98 	& 0.92 	& 0.81 	& 0.94 	& 0.97^* 		& 0.97 & 0.91 & 0.99 & 0.98 \\ 
CookePRE 	& 0.62 	& 0.90^*& 0.47 	& 0.49 	& 0.65 	& 0.90 	& 0.69 	& 0.89 	& 0.70 	& 0.94 	& 0.86 	& 0.94^* 		& 0.90 & 0.96 & 0.97 & 0.95 \\ 
KhademiJOINT& - 	& - 	& 0.74 	& 0.80 	& 0.87 	& 0.53 	& 0.84 	& 0.75 	& 0.90 	& 0.90 	& 0.87 	& 0.96^* 		& 0.96 & 0.91 & 0.97 & 0.95 \\ 
DutchMRG 	& - 	& - 	& 0.19 	& 0.49 	& 0.74 	& 0.72 	& 0.65 	& 0.85 	& 0.82 	& 0.69 	& 0.81 	& 0.92^* 		& 0.77 & 0.75 & 0.87 & 0.91 \\  
DantaleMRG 	& - 	& - 	& 0.72 	& 0.81 	& 0.68 	& 0.76 	& 0.94 	& 0.78 	& 0.96 	& 0.90 	& 0.77 	& 0.82^* 		& 0.72 & 0.86 & 0.89 & 0.85 \\ 
&&&&&&&&&&&& 
\\[-2mm]   
Mean & 0.72 & 0.89 & 0.58 & 0.66 & 0.78 & 0.80 & 0.82 & 0.82 & 0.83 & 0.86 & 0.89 & 0.92 							& 0.85 & 0.88 & 0.91 & 0.92\\
$\mathrm{CI_{low}}$ & 0.65 & 0.85 & 0.53 & 0.61 & 0.74 & 0.76 & 0.79 & 0.79 & 0.80 & 0.82 & 0.86 & 0.90				& 0.83 & 0.86 & 0.89 & 0.89\\
$\mathrm{CI_{high}}$ & 0.78 & 0.92 & 0.63 & 0.70 & 0.81 & 0.83 & 0.85 & 0.85 & 0.86 & 0.89 & 0.91 & 0.93            & 0.87 & 0.90 & 0.93 & 0.93\\
\hline\hline
\end{tabular} 
\end{table*} 

The labels `icra', `sin', `noisex', `noise', 'reverb', `both', `ssn', `cafe', `car', `bottles', `talk', and `ssn' in Figure \ref{fig:scatter} indicate the type of environmental degradation in the data set. The labels `pro' and `un' indicate whether a stimulus was processed by an enhancement algorithm or was unprocessed. The labels `jensen', `hend', `kleijn', `khad', `itfs', `an', and `post' refer to individual data sets within the merged data sets. 

Table \ref{table:tau} displays Kendall's tau coefficient for all data sets and intelligibility metrics and, similarly, Table \ref{table:rho} displays Pearson's correlation coefficient. In both tables, an asterisk is used to indicate when a data set was used during the development of an intelligibility metric. {For the remainder of the paper, 'unseen' refers to a data set that was not used during development, and 'seen' refers to a data set that was used during development.} The mean performance of each intelligibility metric and a confidence interval, [$\mathrm{CI_{low}}$, $\mathrm{CI_{high}}$], with 95\% coverage of the mean performance is also included. The confidence intervals were calculated using the non-parametric $\mathrm{BC_a}$ bootstrap approach \cite{efron1987better}. To do so, 5000 bootstrap sample sequences of $p_c$ and $d(x_c,y_c)$ were generated for each data set and intelligibility metric. The sample distribution of the mean performance of each intelligibility metric was then estimated from the bootstrap sample sequences.

{From here on, subscripts are used to indicate performance criteria for particular intelligibility metrics. For example, $\rho_\mathrm{SIIB}$, refers to the correlation coefficient that SIIB achieved on some data set. }

\subsection{Remarks for the pre-existing metrics}
It is clear that out of the pre-existing metrics SIIB and HASPI have the highest performance overall, on average achieving $\tau_\mathrm{SIIB}$ = 0.79 and $\rho_\mathrm{SIIB}$ = 0.92, and $\tau_\mathrm{HASPI}$ = 0.76 and $\rho_\mathrm{HASPI}$ = 0.89. This performance is followed closely by ESTOI, which has an average score of $\tau_\mathrm{ESTOI}$ = 0.72 and $\rho_\mathrm{ESTOI}$ = 0.86. HEGP has high performance for data sets distorted by additive noise achieving an average score of $\tau_\mathrm{HEGP}$ = 0.73 and $\rho_\mathrm{HEGP}$ = 0.89, but its usefulness is limited to situations where noise is the main source of degradation and where the noise signal is available.

The top performance rating of SIIB may be criticized on the grounds that SIIB has been `over-designed' for the data sets in this evaluation. Although the parameters of SIIB were not intentionally optimized for the data sets in this paper, the developers of SIIB were the only researchers with access to all the data sets and thus had greater opportunity to redesign their algorithm when weaknesses were exposed during SIIBs development.

\begin{figure}[t]
\centering
\includegraphics[width=0.95\columnwidth]{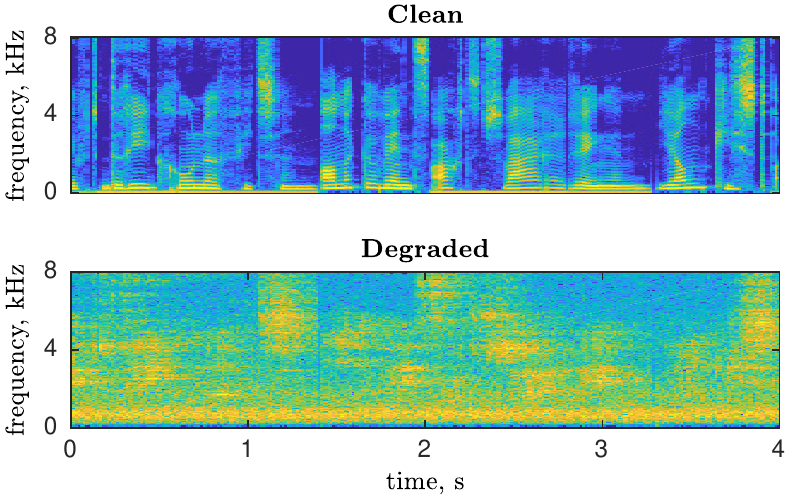}
\vspace*{-0.2cm}
\caption{{An example of a clean and degraded stimulus from HendriksPRE. The severe reverberant distortion 'blurs' the time-alignment between the stimuli.}}
\label{fig:reverb}
\vspace{-0.2cm}
\end{figure}

Many of the intelligibility metrics performed poorly on HendriksPRE. This is likely due to the large T60 time of the room impulse response that causes severe reverberant distortion. {As shown in Figure} \ref{fig:reverb}, the large T60 time somewhat `blurs' the time-alignment of clean and degraded temporal envelopes. Many intrusive intelligibility metrics require that the clean and degraded signals are strictly time-aligned, and thus are over-sensitive to temporal blurring. {Out of all the intelligibility metrics in this evaluation, HASPI achieved the highest performance for HendriksPRE ($\tau_\mathrm{HASPI}$ = 0.78, $\rho_\mathrm{HASPI}$ = 0.92) and is also the only intelligibility metric that included time-alignment processing.} %It is plausible that the time-alignment processing reduces the sensitivity to temporal blurring.}

%It is interesting that SIIB performs well compared to other metrics for JensenMOD ($\tau_\mathrm{SIIB}$ = 0.74, $\rho_\mathrm{SIIB}$ = 0.89) and CookePRE ($\tau_\mathrm{SIIB}$ = 0.76, $\rho_\mathrm{SIIB}$ = 0.94) because these data sets contain noise sources with strong temporal modulations. Generally it has been thought that to predict intelligibility for modulated noise sources, statistics have to be estimated for short-time segments and then combined so that the effect of `listening for glimpses of clean speech' \cite{cooke2006glimpsing, rhebergen2006extended} can be accounted for. Yet SIIB is based on long-term statistics only.

Recall that HASPI is computed as a linear combination of four terms: the cepstral correlation, and three auditory coherence terms. %corresponding to low, mid, and high energy segments. 
The weights in the linear combination were optimized for each data set to maximize performance. None of the other intelligibility metrics modify their parameters based on the data, suggesting that the high performance of HASPI may be attributed to overfitting. To test this hypothesis, HASPI was computed simply by summing the cepstral correlation term and the high-energy auditory coherence term with equal weight. Doing so reduced the mean performance of HASPI to $\tau_\mathrm{HASPI}$ = 0.73 and $\rho_\mathrm{HASPI}$ = 0.88, which is still very high. Thus, the high performance of HASPI is unlikely the result of overfitting.%Moreover, this suggests that if no intelligibility data is available, then a reasonable approach for computing HASPI is to ignore the low-energy and mid-energy coherence terms and give the cepstral correlation and high-energy auditory coherence equal weight.

{Another criteria that can be used to evaluate performance is whether a metric gives consistent predictions across classes of distortions. For example, CookePRE has two distinct classes: stimuli degraded by a competing talker, and stimuli degraded by SSN. Metrics may give consistent intelligibility predictions within a class, but could give inconsistent predictions between classes. %Examples of this can be seen in Figure} \ref{fig:scatter} {for scatter plots corresponding to CookePRE, where the data points tend to form two distinct clusters. 
An example of this can be seen in the scatter plot corresponding to STOI and DutchMRG. STOI gives consistent predictions for JensenPOST, KleijnPRE, and KhademiJOINT, but when the data sets are merged together we see distinct clusters corresponding to each data set. %Recall that the listening tests for the data sets in DutchMRG were conducted by the same laboratory group with the same speech corpus. 
This means that for a given clean stimulus, a STOI score of 0.5 for noise-reduced speech and a STOI score of 0.5 for pre-processed speech could correspond to different intelligibility scores. }

\begin{table*}[t!]
\centering
\renewcommand{\arraystretch}{1.2}
\caption{\label{table:seen} {Mean performance of pre-existing intelligibility metrics for 'seen' and 'unseen' data sets.}}
\begin{tabular}{l dddddddddddd}  
\hline\hline
& \multicolumn{1}{c}{\text{SII}} & \multicolumn{1}{c}{\text{HEGP}} & \multicolumn{1}{c}{\text{NCM-BIF}} & \multicolumn{1}{c}{\text{QSTI}} & \multicolumn{1}{c}{\text{CSII-MID}} & \multicolumn{1}{c}{\text{MIKNN}} & \multicolumn{1}{c}{\text{SIMI}} & \multicolumn{1}{c}{\text{$\mathrm{sEPSM^{corr}}$}} & \multicolumn{1}{c}{\text{STOI}} & \multicolumn{1}{c}{\text{ESTOI}} & \multicolumn{1}{c}{\text{HASPI}} & \multicolumn{1}{c}{\text{SIIB}} 
\\        
\hline
mean $\tau^\mathrm{seen}$ & - & 0.72 & 0.67 & 0.69 & - & 0.77 & 0.84 & 0.70 & 0.87 & 0.78 & - & 0.79 \\                 
mean $\tau^\mathrm{unseen}$ & 0.57 & 0.73 & 0.46 & 0.45 & 0.63 & 0.63 & 0.60 & 0.66 & 0.66 & 0.69 & 0.76 & - \\ 
\\[-3.5mm] \hline \\[-3.5mm]
mean $\rho^\mathrm{seen}$ & - & 0.90 & 0.89 & 0.86 & - & 0.92 & 0.95 & 0.84 & 0.97 & 0.93 & - & 0.92 \\                 
mean $\rho^\mathrm{unseen}$ & 0.72 & 0.88 & 0.56 & 0.60 & 0.78 & 0.78 & 0.78 & 0.82 & 0.80 & 0.84 & 0.89 & - \\
\hline\hline
\end{tabular} 
\end{table*} 

\subsection{{Investigating the performance in terms of generalization}}
Considering only entries in Table \ref{table:tau} and Table \ref{table:rho} that have an asterisk, the mean performance of all such entries for all pre-existing metrics and data sets is $\tau$ = 0.78 and $\rho$ = 0.92. Considering only entries that do not have an asterisk, the mean performance for all pre-existing metrics and data sets is $\tau$ = 0.62 and $\rho$ = 0.76. This result demonstrates that, in general, intelligibility metrics have high performance for seen data sets, and poor performance for unseen data sets.

To further investigate the performance of intelligibility metrics in terms of their ability to generalize, Table \ref{table:seen} displays the mean performance for unseen data sets and seen data sets for each pre-existing intelligibility metric. HASPI has the highest performance for unseen data sets achieving $\tau_\mathrm{HASPI}^\mathrm{unseen}$ = 0.76 and $\rho_\mathrm{HASPI}^\mathrm{unseen}$ = 0.89. HEGP also has high performance for unseen data sets, however, recall that HEGP was evaluated exclusively on data sets with additive noise degradation.

STOI and SIMI both have outstanding performance for seen data sets ($\tau_\mathrm{STOI}^\mathrm{seen}$ = 0.87, $\rho_\mathrm{STOI}^\mathrm{seen}$ = 0.97, and $\tau_\mathrm{SIMI}^\mathrm{seen}$ = 0.84, $\rho_\mathrm{SIMI}^\mathrm{seen}$ = 0.95), but poor performance for unseen data sets ($\tau_\mathrm{STOI}^\mathrm{unseen}$ = 0.66, $\rho_\mathrm{STOI}^\mathrm{unseen}$ = 0.80, and $\tau_\mathrm{SIMI}^\mathrm{unseen}$ = 0.60, $\rho_\mathrm{SIMI}^\mathrm{unseen}$ = 0.78). This is because STOI and SIMI were specifically designed for speech processed by ITFS and noise-reduction algorithms, whereas the data sets in this evaluation include degradation caused by reverberation and modulated noise sources. Similarly, NCM-BIF was designed specifically for speech processed by noise-reduction algorithms. Observe that in Figure \ref{fig:scatter} NCM-BIF has good performance for the data sets with noise-reduction: HuPOST, JensenPOST, and TaalPOST, but poor performance for the remaining data sets. These results show the danger of using intelligibility metrics outside of their intended domain.

In light of the above paragraphs, to ensure that future intelligibility metrics generalize to new data sets and give consistent predictions between classes, it may be more beneficial to gather data points with different types of degradation than to collect many data points for a single type of degradation. This notion is consistent with the high performance of HASPI, which considered six types of degradation during development: additive noise, envelope-clipping, ITFS processing, frequency-compression, noise reduction, and vocoded-speech.%In contrast, during the development of STOI, only two types of degradation were considered: ITFS processing and noise-reduction, and during the development of NCM-BIF only one type of degradation was considered: noise-reduction.

\subsection{{Remarks for the modified intelligibility metrics}}
In general, removing the KLT from SIIB significantly reduced performance (on average $\tau_\mathrm{SIIB^{no\_KLT}}= 0.69$ and $\rho_\mathrm{SIIB^{no\_KLT}}= 0.85$). Furthermore, introducing the KLT to STOI improved performance (on average $\tau_\mathrm{STOI^{KLT}} = 0.73$ and $\rho_\mathrm{STOI^{KLT}} = 0.88$). The increase in overall performance for $\text{STOI}^\text{KLT}$ is mainly due to large increases in performance for JensenMOD, HendriksPRE, and CookePRE. Note that $\text{STOI}^\text{KLT}$ performs worse than STOI for KjemsITFS and TaalPOST, however, these are the same data sets that were used to tune the parameters of STOI during STOIs development. 

The five intelligibility metrics with the highest performance: SIIB, $\text{SIIB}^\text{Gauss}$, $\text{STOI}^\text{KLT}_\text{gamma}$, HASPI, and $\text{STOI}^\text{KLT}$ are also the only metrics that decorrelate log-spectra. This outcome clearly demonstrates the advantage that can be obtained by reducing the statistical dependencies between input features.

Recall that ESTOI was proposed as an extension to STOI that can 'listen to glimpses of clean speech'. Interestingly, for the data sets that contain modulated noise, $\text{STOI}^\text{KLT}$ has similar performance to ESTOI (for JensenMOD, $\tau_\mathrm{STOI^{KLT}}=0.72$, $\rho_\mathrm{STOI^{KLT}}=0.90$, and for CookePRE, $\tau_\mathrm{STOI^{KLT}}=0.87$, $\rho_\mathrm{STOI^{KLT}}=0.96$). SIIB and $\text{SIIB}^\text{Gauss}$, which are based on long-term statistics, also have good performance for JensenMOD and CookePRE. Such results contest the idea that short-time segmentation is necessary for predicting the intelligibility of modulated noise sources. 

On average $\text{STOI}^\text{KLT}_\text{gamma}$ achieved $\tau_\mathrm{STOI^{KLT}_{gamma}} = 0.76$ and $\rho_\mathrm{STOI^{KLT}_{gamma}} = 0.91$. Thus, by introducing the KLT to STOI and using a more realistic auditory model, performance competitive with SIIB could be obtained. This means that for some representations of speech signals, the correlation coefficient and the KNN mutual information estimator can quantify distortion equally well. A partial explanation for this result can be found by considering the high performance of $\text{SIIB}^\text{Gauss}$ ($\rho_\mathrm{SIIB^{Gauss}} = 0.92$ and $\tau_\mathrm{SIIB^{Gauss}} = 0.79$), which suggests that the Gaussian communication channel is a reasonable approximation of the true communication channel for many real-word distortions.

Finally, recall that $\text{SIIB}^\text{Gauss} = -\frac{F}{2K}\sum_j \log_2(1-r^2\rho_j^2)$. Since $r$ and $\rho_j$ are between -1 and 1, the product of their squares is likely to be small, particularly for challenging listening environments. Using the approximation $\log_2(1+a)\approx a/\ln(2)$ for small $a$, we have that $\text{SIIB}^\text{Gauss} \approx \frac{F}{2K\ln(2)} r^2 \sum_j \rho^2_j$. This approximation strongly resembles the distortion measure used by $\text{STOI}^\text{KLT}$ and $\text{STOI}^\text{KLT}_\text{gamma}$, which can be written as $\sum_j \sum_t\rho_{j,t}$, where $t$ is the short-time segment index. %This explains why $\text{STOI}^\text{KLT}_\text{gamma}$ has performance similar to $\text{SIIB}^\text{Gauss}$. %The effect of using the correlation coefficient instead of the square of the correlation coefficient is that the relative importance of poorly correlated channels is over-weighted. Recall that the original STOI algorithm includes a clipping procedure to ensure that the SDR does not fall below -15 dB. The clipping procedure mitigates the effect of over-weighting poorly correlated channels.

\section{\label{sec:conclusions}Conclusions}
In this paper, the accuracy of 12 intelligibility metrics from the literature was evaluated using the results of 11 listening tests. The stimuli included pre-processing enhancement, post-processing enhancement, and environmental distortions such as noise and reverberation. {In order to analyze why the top performing metrics have high performance, four new intelligibility metrics were proposed.} The main conclusions are as follows.
\begin{enumerate}
\item {Out of the pre-existing metrics, SIIB and HASPI had the highest overall performance.} 

\item {Many intrusive metrics struggle with severe reverberant distortion. This may be because they are over-sensitive to the time-alignment of clean and distorted temporal envelopes.}

\item {In general, intelligibility metrics perform more poorly on unseen data sets than on seen data sets. For this reason, caution should be taken when using intelligibility metrics outside of their intended domain.}

\item {For unseen data sets, HASPI had the highest performance. This suggests that HASPI is appropriate for situations where many types of potentially new speech material and distortions are likely. Additionally, unlike the other metrics, HASPI has built-in time-alignment processing and can account for hearing impairments.}

\item {The five intelligibility metrics with the highest overall performance are also the only metrics that decorrelate log-spectra. On average, introducing the KLT to STOI improved performance and removing the KLT from SIIB reduced performance. These results demonstrate the advantage of removing statistical dependencies between input features.}

\item {The high performance of $\text{SIIB}^\text{Gauss}$ suggests that the Gaussian communication channel is a reasonable approximation of the true communication channel for many real-world distortions. Additionally, $\text{SIIB}^\text{Gauss}$ has performance similar to SIIB, but takes less time to compute by two orders of magnitude.\footnote{{MATLAB implementations of $\text{SIIB}^\text{Gauss}$ and SIIB are available at www.stevenvankuyk.com/MATLAB\_code}}}
%\wbk{footnotes go after punctuation!}
%\wbk{But what is the effect of not stacking and no production noise on the rate?}

\item {It was shown that $\text{STOI}^\text{KLT}$ and $\text{STOI}^\text{KLT}_\text{gamma}$ can be interpreted as approximations of $\text{SIIB}^\text{Gauss}$.}

\end{enumerate}  

\section*{Acknowledgments}
The authors would like to thank the following researchers for providing intelligibility data and MATLAB implementations of their intelligibility metrics: Asger Andersen, Fei Chen, Martin Cooke, Jesper Jensen, James Kates, Helia Relano-Iborra, Jo\~{a}o Santos, and Yan Tang. The authors would also like to thank Ulrik Kjems, Kuldip Paliwal, Jalal Taghia, and Cees Taal, for making their materials publicly available. {Finally, the authors would like to thank the three anonymous reviewers for their insightful comments.}

\ifCLASSOPTIONcaptionsoff
  \newpage
\fi

%\vfill
%\newpage


\bibliographystyle{IEEEtran}
%\bibliography{citations}


%\begin{thebibliography}{1}
%\end{thebibliography}


\vfill
\pagebreak
\begin{IEEEbiography}[{\includegraphics[width=1in,height=1.25in,clip,keepaspectratio]{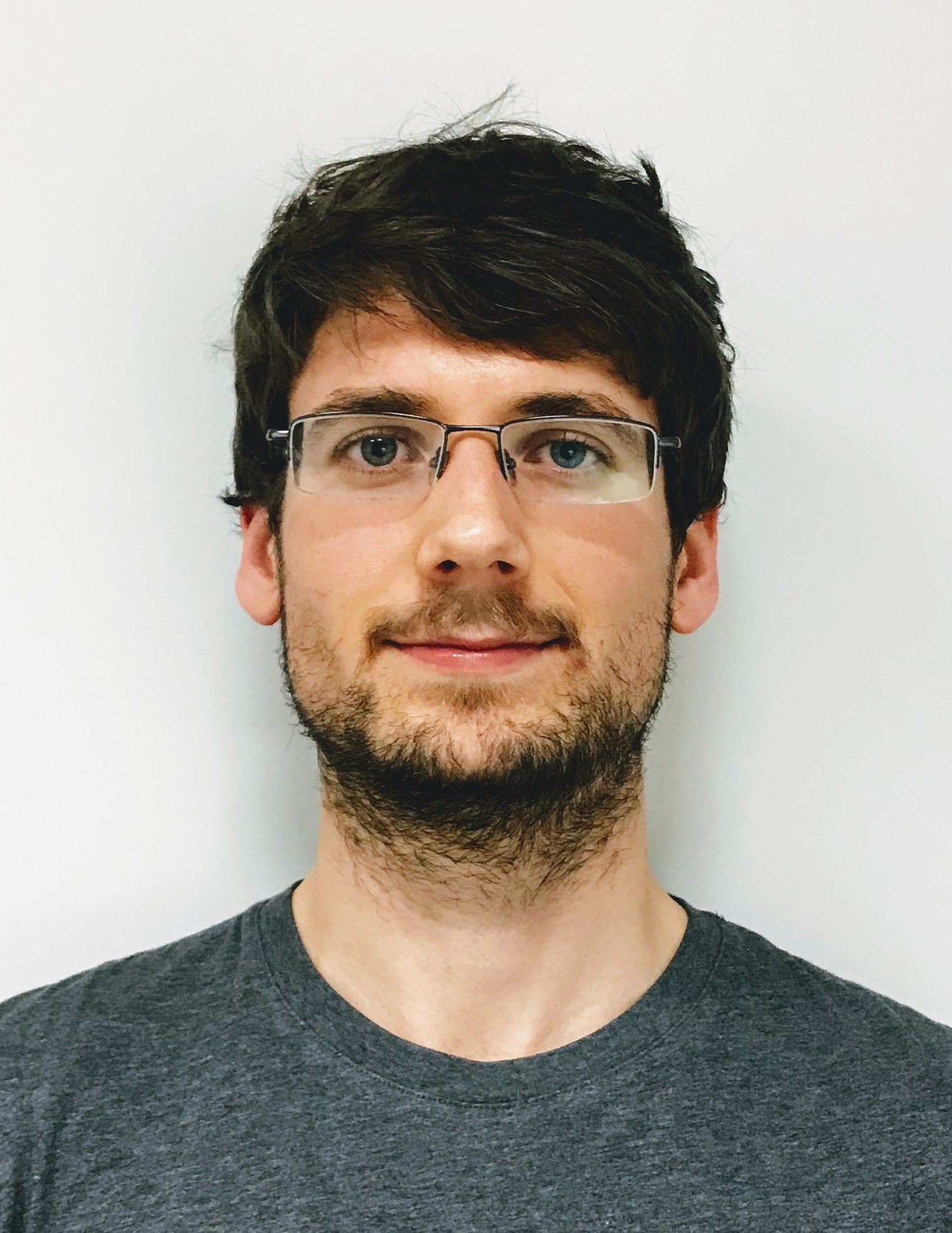}}]{Steven Van Kuyk}
received the BEHONS degree in electronic and computer systems engineering from Victoria University of Wellington, Wellington, New Zealand in 2015, and is currently working toward a Ph.D. degree at Victoria University of Wellington, Wellington, New Zealand. In 2014 he worked at Fisher \& Paykel Healthcare, Auckland, New Zealand, and in 2017 he worked at Apple Inc., Cupertino, U.S.A. His research interests include signal processing, information theory, and machine learning, for applications in audio processing.
\end{IEEEbiography}

\begin{IEEEbiography}[{\includegraphics[width=1in,height=1.25in,clip,keepaspectratio]{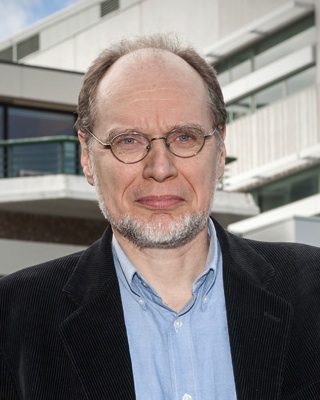}}]{W. Bastiaan Kleijn}
W. Bastiaan Kleijn is Professor at Victoria University of Wellington, New
Zealand and Professor (part-time) at Delft University of Technology (DUT). 
He was Professor and Head of the Sound and Image Processing Laboratory at
KTH in Stockholm, 1996-2010. Kleijn was a founder of Global IP Solutions, a
company that provided the enabling audio technology to Skype. It was acquired
by Google in 2010.  He has served on a number of editorial Boards including 
those of the IEEE Trans. Audio Speech Language Processing, Signal Processing,
IEEE Signal Processing Letters, and IEEE Signal Processing Magazine. He was the
Technical Chair of ICASSP 1999 and EUSIPCO 2010, and two IEEE workshops. Kleijn
received a PhD degree in Electrical Eng. from TU Delft, a PhD in Soil Science
and an MSc in Physics from the University of California, Riverside, and an MSEE
from Stanford University. He is Fellow of the IEEE since 1999.
\end{IEEEbiography}

\begin{IEEEbiography}[{\includegraphics[width=1in,height=1.25in,clip,keepaspectratio]{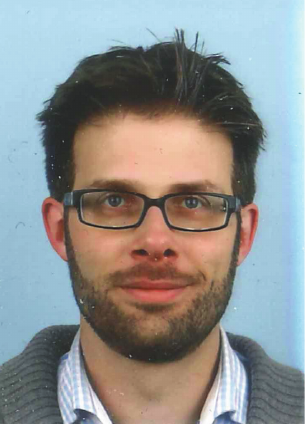}}]{Richard C. Hendriks}
Richard Christian Hendriks was born in Schiedam, The Netherlands. He received the B.Sc., M.Sc. (cum laude), and Ph.D. (cum laude) degrees in electrical engineering from the Delft University of Technology, Delft, The Netherlands, in 2001, 2003, and 2008, respectively. He is currently an Associate Professor in the Circuits and Systems (CAS) Group, Faculty of Electrical Engineering, Mathematics and Computer Science, Delft University of Technology. His main research interest is on biomedical signal processing, and, audio and speech processing, including speech enhancement, speech intelligibility improvement and intelligibility modelling. In March 2010, he received the prestigious VENI grant for his proposal “Intelligibility Enhancement for Speech Communication Systems”. He obtained several best paper awards, among which the IEEE Signal Processing Society best paper award in 2016. He is an Associate Editor for the IEEE/ACM Trans. on Audio, Speech, and Language Processing and the EURASIP Journal on Advances in Signal Processing.
\end{IEEEbiography}

\vfill
%\end{IEEEbiography}
%% insert where needed to balance the two columns on the last page with
%%\newpage


%\vfill


\end{document}